\documentclass[useAMS,usenatbib]{mn2e}  
\usepackage{amssymb,amsmath}
\usepackage{amsfonts}
\usepackage{graphicx}
\usepackage{booktabs}
\usepackage{tabularx}

\numberwithin{equation}{section}

\def\d{{\rm d}}
\def\e{{\rm e}}
\newcommand{\p}{\partial}

\newcommand{\suml}{\sum\limits}
\newcommand{\il}{\int\limits}
\renewcommand{\i}{\mbox{i}}

\renewcommand{\[}{\left[}

\renewcommand{\Im}{\textrm{Im}\,}

\begin{document}

\title[The linear eigenvalue problem]{The linear eigenvalue problem for barotropic selfgravitating discs}
\author[E. V. Polyachenko]{E.~V.~Polyachenko, \thanks{E-mail: epolyach@inasan.ru}\\
Institute of Astronomy, Russian Academy of Sciences, 48 Pyatnitskya St., Moscow 119017, Russia}



\maketitle	

\begin{abstract}

Gaseous rotating razor-thin discs are a testing ground for theories of spiral structure that try to explain appearance and diversity of disc galaxy patterns. These patterns are believed to arise spontaneously under the action of gravitational instability, but calculations of its characteristics in the gas are mostly obscured, presumably due to a difficult outer boundary condition. The paper suggests a new effective method for finding the spiral patterns based on an expansion of small amplitude perturbations over finite radial elements. The final matrix equation is extracted from the original hydrodynamical equations without the use of an approximate theory and has a form of the linear algebraic eigenvalue problem. The method is applied to an exactly solvable model with finite outer boundary and to a galactic disc model.

\end{abstract}


\section{Introduction} 

Over years, quasi-stationary spiral structure hypothesis (QSSS) has been used for explaining formation and shapes of spirals in disc galaxies. The spirals are treated there as density waves, possibly arising spontaneously in an axisymmetric background due to gravitational instability. Mathematically speaking, it is the eigenvalue problem where eigenvalues stand for pattern speeds and growth rates, and eigenfunctions describe spiral patterns.
  
The galactic disc is a complicated multi-component system interacting with its surroundings such as bulge, stellar halo, dark matter, etc., so any realistic setting of this problem is not yet possible. There hoping was that the essence of the phenomena can be grasped by studying the most important components, first of all, a thin stellar disc. The eigenvalue problem for razor-thin stellar discs was pioneered by \citet{K71, K77}, who developed a matrix method, nonlinear with respect to eigenvalues. The complexity of the method, however, prevented it from widespread use. 

\citet{EP05} suggests another method which takes advantage of the standard linear algebraic eigenvalue problem for calculation of stellar disc eigenmodes, and a whole spectrum of unstable modes can be obtained at once. \citet{J10} continued to explore the linear eigenvalue problem, employing Bubnov-Galerkin finite element method (FEM). A comparison of linear methods can be found in \citet{PJ15}.

Stellar and gaseous discs are equivalent in the cold limit, and many aspects of stellar disc dynamics can be understood in the hydrodynamical approach. This has been the rationale for the development of the theory of gaseous discs as a substitution for stellar discs \citep[see, e.g.,][]{B14}. The gaseous approach is less complicated since it includes Euler equation and the barotropic equation of state instead of more detailed kinetic description for stars. However, propagation of density waves across Lindblad resonances is different in these media \citep{BT08}, which results in a specific boundary condition for gaseous discs. A great deal of attention has been paid by C.C.Lin and co-authors to address this problem carefully, and the recipes for eigenmode's calculation were summarised in \citet{P83}. In his work, the solution of the eigenvalue problem is constructed in two regions: a numerical solution in the inner region, and the analytical approximate WKB solution in the outer region.   

To a large extent, moderate success of QSSS is explained by the uncertainty of the initial axisymmetric configurations. So, an importance of a generating mechanism of spiral structures could be determined only statistically, by analysing a diversity of initial states. In this situation speed and accuracy of eigenmode's calculation are crucial. The goal of this paper is to suggest and test an effective matrix method applicable to various gaseous discs using the Galerkin formulation of one dimensional FEM.

The structure of the paper is the following. In Section 2 we derive a matrix form of the linearised hydrodynamical equations. Section 3 is devoted to a disc with finite outer boundary and Section 4 describes the cored exponential disc model in two ways: with softened gravity and no pressure, and gaseous disc with pressure. The last Section 5 contains summary and outlines perspectives.

\section{Linearised equations in the finite element form}

An axisymmetric gaseous razor-thin disc can be defined by surface density $\Sigma_0(R)$, sound speed $c_\textrm{s}(R)$, and angular velocity $\Omega(R)$. In equilibrium, it meets the following condition:
\begin{equation}
   R\Omega^2(R) = \frac{\d }{\d R}  \left[ \Phi_0(R) + h_0(R) \right]\ ,
   \label{eq:eqd}
\end{equation} 
where $\Phi_0(R)$ is the potential produced by the disc and an external source, $h_0 = \int (c_\textrm{s}^2/\Sigma_0) \d \Sigma_0$ is the enthalpy.

Stability analysis prescribes an ansatz for all perturbed functions in the form $f(R) \e^{\i (m\theta-\omega t)}$, where $f(R)$ is the  amplitude, $\theta$ is the azimuth, $m$ is the azimuthal number, $\omega \equiv m\Omega_\textrm{p} + \i\omega_\textrm{I}$ is a frequency of the perturbation. Unstable solutions correspond to $\omega_\textrm{I} > 0$. From the linearised Euler equation one has \citep[e.g.,][]{FP84}:
\begin{align}
   &-\i \omega_* v_{R}  - 2\Omega v_{\theta}  = - \frac{\d }{\d R} (\Phi + h)\ , \label{eq:vr}\\
   &-\i \omega_* v_{\theta} + \frac{\kappa^2}{2\Omega} v_{R} = -\frac{\i m}{R} (\Phi+h)\ , \label{eq:vt}
\end{align}
where $\omega_* \equiv \omega - m\Omega$,  $\kappa$ is the epicyclic frequency:
\begin{equation}
   \kappa^2 \equiv 4\Omega^2 + R\frac{\d \Omega^2}{\d R} \ ,
\end{equation} 
$v_{R}, v_{\theta}$, $h$, $\Sigma$, $\Phi$ are the amplitudes of perturbations for velocity components, enthalpy, surface density, and potential.

The linearised barotropic equation of state and continuity equation give, respectively:
\begin{align}
   &h = c_\textrm{s}^2 \frac{\Sigma}{\Sigma_0} \label{eq:h}\ , \\
   &-\i \omega_* \Sigma + \frac1R \frac{\d}{\d R} (R \Sigma_0 v_{R}) + \frac{\i m \Sigma_0}R v_{\theta} = 0 \ .\label{eq:sigma}
\end{align}

It is known that softened gravity can mimic the effect of velocity dispersion \citep{M71, E74, T77}. In order to be able to examine softened gravity discs, we assume that the potential is related to the surface density as
\begin{equation}
  \Phi(R) = \il_0^{\infty} \d R' \, R' G_m (R,R') \Sigma (R') \equiv \hat G_m \Sigma\ , \label{eq:phi}
\end{equation}
where $\hat G_m$ denotes a new operator for integration with the Green function,
\begin{equation}
  G_m (R, R') = -G\il_{-\pi}^{\pi} \d \theta \frac{\cos(m\theta)}{[ R^2 + {R'}^2 - 2R R' \cos\theta + b^2 ]^{1/2} }\ , \label{eq:green}
\end{equation}
and $b$ is the softened gravity parameter. Introducing an auxiliary function  
\begin{equation}
  f_m (z) \equiv \il_{-\pi}^{\pi} \d \theta \frac{\cos(m\theta)}{[1 - (1-z) \cos\theta ]^{1/2} }\ , \label{eq:f_green}
\end{equation}
the Green function can be written as follows:
\begin{equation}
   G_m (R, R') = \frac{-1}{[R^2 + {R'}^2+ b^2]^{1/2}}  f_m \left( \frac {(R- R')^2+ b^2}{R^2 + {R'}^2 + b^2}\right)\ .
 \label{eq:fm}
\end{equation}

For discs without softening, the auxiliary function has a weak singularity at $z=0$. In particular,
\begin{align}
  f_0(z)  &= \sqrt{2} \ln \frac{32}{z} + O(z\ln z)\ , \\
  f_1(z)  &= f_0(z) - 4\sqrt{2} + O(z\ln z)\ , \\
  f_2(z)  &= f_0(z) - \frac{16}3 \sqrt{2} + O(z\ln z)\ .
\end{align}

The desired set of integro-differential equations is:
\begin{align}
   \omega v_{R}       &= m\Omega v_{R} + 2\i \Omega v_{\theta}  - \i \frac{\d }{\d R} \left[ \hat G_m  \Sigma  + \frac{c_\textrm{s}^2}{\Sigma_0} \Sigma \right] \ , \label{eq:vr1}\\
   \omega v_{\theta}  &= -\i \frac{\kappa^2}{2\Omega} v_{R} + m\Omega v_{\theta} + \frac{m}{R} \left[ \hat G_m  \Sigma  + \frac{c_\textrm{s}^2}{\Sigma_0} \Sigma \right] \ , \label{eq:vt1} \\
   \omega \Sigma      &= -\frac{\i}R \frac{\d}{\d R} \left( R \Sigma_0 v_{R} \right) + \frac{m \Sigma_0}R v_{\theta} + m\Omega \Sigma\ .\label{eq:sigma1}
\end{align}
Now the standard route is to express the velocity components in favour of $\Sigma$ and solve eq. (\ref{eq:sigma1}) with $\omega$ as a parameter determined from appropriate boundary conditions. 

In the finite element method, the domain is divided to several `elements'. In 1D case, this is done by defining the radial nodes, $R_{\textrm{in}} = R_0 < R_1 < \dots < R_{N-1} < R_N = R_{\textrm{out}}$, where $N$ is the number of elements. Afterwards, basis functions  $\phi_j(R)$ should be defined. The simplest ones have a form of a triangle having a value of 1 at one node and zero in adjacent nodes. Basis functions of higher degree of smoothness are constructed using Lagrange polynomials, but then additional internal nodes are required. Denoting a degree of the polynomial by $N_\textrm{d}$, the number of internal nodes is $N(N_\textrm{d}-1)$, and the total number of nodes is $N_\textrm{t} \equiv N_\textrm{d}N+1$. 

An unknown function $f$ is approximated by a linear combination:
\begin{equation}
  f(R) \approx \suml_{j=0}^{N_\textrm{t}} F_j\,\phi_j\ ,
  \label{eq:f}
\end{equation}
where $F_j$ is the value of the function in node $R_j$.

Our goal is to obtain the linear algebraic eigenvalue problem in the form:
\begin{equation}
	\omega \mathbf{x} = \mathbf{A} \mathbf{x}\ ,
	\label{eq:me}
\end{equation}
where $\mathbf{A}$ is a matrix, and $\mathbf{x}$ is a vector. Let $v_R$, $v_\theta$ and relative surface density $\eta \equiv \Sigma/\Sigma_0$ be independent unknown functions, and $U_j$, $V_j$, $S_j$ ($j=0, ..., N_\textrm{t}$) -- corresponding expansion coefficients. Vector $\mathbf{x}$ is constructed by concatenation:
\begin{equation}
	\mathbf{x} \equiv [U_0, ..., U_{N_\textrm{t}}, V_0, ..., V_{N_\textrm{t}}, S_0, ..., S_{N_\textrm{t}}]^{T}
\end{equation}
($T$ denotes matrix transposition). 

The weak matrix form is obtained by multiplying eqs.\,(\ref{eq:vr1}. \ref{eq:vt1}) to  $R\mu\phi_k(R)$, and eq.\,(\ref{eq:sigma1}) to  $R\nu\phi_k(R)$, $k=0, ..., N_\textrm{t}$, and integrating over radius. For the left-hand sides, one obtains:
\begin{equation}
	\mathbf{M} \mathbf{x} \equiv \left(
	  \begin{array}{lll}
	  	(\mu R)_{kj}  & 0  &  0 \\[3mm]
	  	0  & (\mu R)_{kj}   & 0 \\[3mm]
		0 & 0 & (\nu R)_{kj}
	   \end{array} \right)\mathbf{x}\ .
	   \label{eq:M}
\end{equation}
The right-hand sides give $ \mathbf{L} \mathbf{x}$, where
\begin{equation}
	\mathbf{L} \equiv \left(
	  \begin{array}{lll}  
	  	m (R\mu \Omega)_{kj}  & 2\i (R\mu \Omega)_{kj}  &  -\i P'_{kj} \\[3mm]
	  	-\i \left( \displaystyle \frac{R\mu \kappa^2}{2\Omega} \right)_{kj}  & m (R\mu \Omega)_{kj}     &  m P_{kj} \\[3mm]
		-\i D_{kj} & m (\nu)_{kj}  &  m (R\nu\Omega)_{kj}   
	   \end{array} \right).
	    \label{eq:L}
\end{equation}
In the matrices, $(...)_{kj}$  denote $\int \d R\, (...) \phi_k(R) \phi_j(R)$, 
\begin{multline}
P_{kj} = \int \d R \d R' \, \mu(R) \phi_k(R) G_m(R,R') R' \Sigma_0 (R') \phi_j(R')    \\
+ \int \d R \, \mu(R) \phi_k(R) c_\textrm{s}^2(R) \phi_j(R) \ ,
\label{eq:p}
\end{multline}
\begin{multline}
P'_{kj} = \int \d R \d R' \, \mu(R) \phi_k(R) R \frac{\d G_m(R,R') }{\d R}  \Sigma_0 (R') \phi_j(R')  \\
+ \int \d R \, \mu(R) \phi_k(R)  R \frac{\d }{\d R} \Big[ c_\textrm{s}^2(R) \phi _j(R) \Big]\ ,
\label{eq:ps}
\end{multline}
\begin{equation}
D_{kj} = \int \d R \, \nu(R) \frac{\phi_k(R)}{\Sigma_0 (R)} \frac{\d} {\d R} \Big[ R \Sigma_0(R) \phi_j(R) \Big]\ .
\end{equation}
Ranks of the matrices $ \mathbf{M}$ and $ \mathbf{L}$ equal to $3N_\textrm{t}$. The desired linear form (\ref{eq:me}) is obtained if $ \mathbf{A} \equiv \mathbf{M}^{-1}  \mathbf{L}$. 

Calculation of integrals involving $G_m(R,R')$ is the most time-consuming since all of them are non-zero. Besides in case $b=0$, the main complication is the singularity of the integrands when $R'$ is close to $R_i$. To handle this singularity, we divide $f_m(z)$ in (\ref{eq:fm}) into regular and singular parts:
\begin{equation}
	f(z) = \overline{f}(z) - \sqrt{2} \ln[ (R'-R)^ 2+ b^2 ]\ .
\end{equation}
Integrals containing $\overline{f}(z)$ are integrated in the usual way. To evaluate the singular contribution to the integral from the interval $[R_n, R_{n+1}]$ we apply a four-point quadrature rule of the form 
\begin{equation}
\int\limits_{R_{n}}^{R_{n+1}} dR' w_b(R-R') g(R') = \sum\limits_{j=0}^3 W^j_b g_{3n+j}\ ,
\label{eq:iw}
\end{equation}
where $g$ is a smooth function, the weight function
\begin{equation}
w_b(x) = \ln[ x^ 2+ b^2 ]\ , 
\label{eq:wf}
\end{equation}
and $g_{3n+j} \equiv g(R_n + j(R_{n+1}-R_{n})/3)$. The weights $W^j_b$ are chosen so that the integration is exact if $g(R)$ is a cubic polynomial, and can be evaluated analytically using the integrals 
\begin{equation}
\int_a^b {R'}^n w_b(R-R') \d R'\ , \quad n=0, ..., 3
\end{equation}
\citep{Press92, T01}.

Despite the original hydrodynamical eigenvalue problem requires boundary conditions, its matrix counterpart essentially has none. Our strategy is to apply the matrix equation as it is, and to filter out the solutions with needed physical properties. A so-called {\it regularity} boundary condition, which is vanishing of the perturbation amplitudes in the centre $\propto R^m$, can be applied by taking to zero the expansion coefficients for $j=0$. In the remaining sections, we shall consider bisymmetric ($m=2$) modes only.

The weak formulation allows flexibility in choosing of the weight functions $\mu(R)$ and $\nu(R)$. Calculations below assume $\mu=\nu=1$ and $\mu=\nu=\Sigma_0(R)$ and show that spectra are insensitive to the choice of the functions when applied to models without acoustic waves allowed to go to infinity (the first two models considered below). However, in the last model, the modes with low growth rates can be sensitive to changing of these functions and smoothness of the basis functions.

\section{Hunter-Shukhman model with uniform rotation}

\citet{H63} suggested a cold exactly solvable model with a uniform law of rotation, which is produced by a disc of finite radius $a$ and surface density law
\begin{equation}
 \Sigma_0 (R) = \Sigma_* \xi\ ,\quad \xi \equiv \Big[1 - R^ 2/a^2\Big]^{1/2}\ .
 \label{eq:sd_rigid}
\end{equation}
The self-consistency condition leads to the harmonic potential $\Phi_0(R) = \Omega^2_\textrm{D} R^2/2$ with 
\begin{equation}
 \Omega^2_\textrm{D}  =  \frac{\pi^2 G\Sigma_*}{2a}\ .
 \label{eq:om_rigid}
\end{equation}

I.\,G.\,Shukhman (unpublished) has generalised the model by including pressure with polytropic index $\gamma=3$:
\begin{equation}
p_0 (R) = \frac 13 \Sigma_* c_*^2 \xi^3 \ ,
 \label{eq:p_rigid}
\end{equation}
leading to sound speed $c_\textrm{s} =  c_* \xi$, and the following condition for radial equilibrium with constant angular velocity $\Omega$:
\begin{equation}
 \Omega^2_\textrm{D} =  \Omega^2 + \frac{c_*^2}{a^2} \ .
 \label{eq:eq_rigid}
\end{equation}
Eigenfunctions of the surface density and potential coincide with ones of the cold case:
\begin{align}
 \Sigma_n(R) &= \Sigma_* \frac{P_n^m(\xi)}{\xi} \ ,  \label{eq:efs_rigid} \\
 \Phi_n(R) &= - 4\Gamma_n^m a^2  \Omega^2_\textrm{D}  P_n^m(\xi) \ ,
 \label{eq:efp_rigid}
\end{align}
where $P_n^m$ are the associated Legendre polynomials, and
\begin{equation}
 \Gamma_n^m  =  \frac{(n+m)! (n-m)!} {\displaystyle 2^{2n+1} \left[\left(\frac{n+m}{2}\right)!  \left(\frac{n-m}{2}\right)! \right]^2  }\ ,\quad n=2,4,6,...
 \label{eq:gm_rigid}
\end{equation}
Note that $\Sigma_n(R)$ has an integrable singularity, which is an artefact of the Eulerian description which assumes that the edge of the disc is fixed. To get rid of the singularity, we changed the relative surface density $\eta$ to $\tilde\Sigma \equiv \Sigma \xi$ that should be simply Legendre polynomials.

For non-zero pressure, gravitational perturbation $\Phi_n$ is substituted by $\Phi_n + c_\textrm{s}^2 \Sigma_n/\Sigma_0$. Thus, in Hunter characteristic equation the term $- 4\Gamma_n^m a^2  \Omega^2_\textrm{D}$ is changed to $c_*^2- 4\Gamma_n^m a^2  \Omega^2_\textrm{D}$, which results in the modified characteristic equation:
\begin{equation}
 1 = \frac{c_*^2/ a^2 - 4\Gamma_n^m   \Omega^2_\textrm{D}}{\omega_*^2 - 4\Omega^2} \left( n^2 + n - m^2 - \frac{2m}{\omega_*} \right)\ .
 \label{eq:dr_rigid}
\end{equation}
In the limit of large $n$, $\Gamma_n^m \approx 1/(\pi n)$, and (\ref{eq:dr_rigid}) reduces to the WKB dispersion relation:
\begin{equation}
\omega_*^2 - 4\Omega^2 = c_*^2\frac{n^2} {a^2} - \frac{4n}{\pi } \Omega^2_\textrm{D}  \ .
 \label{eq:wkb_rigid}
\end{equation}
Toomre stability criterion for this model reads:
\begin{equation}
Q = \frac{\kappa c_\textrm{s}}{\pi G \Sigma_0} =  \frac{\pi \Omega c_*}{a\Omega^2_D} > 1 \ .
 \label{eq:Q_rigid}
\end{equation}

\begin{figure*}
  \centerline{\includegraphics [width = 80mm]{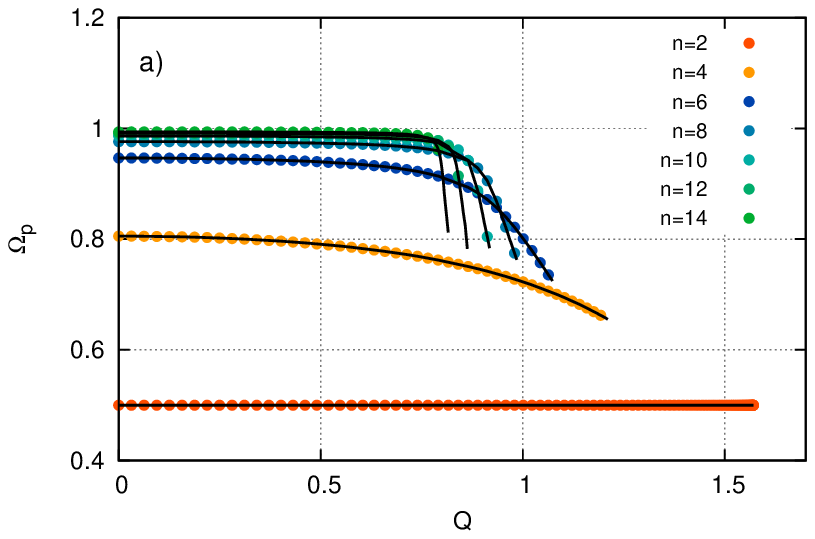}     \includegraphics [width = 80mm]{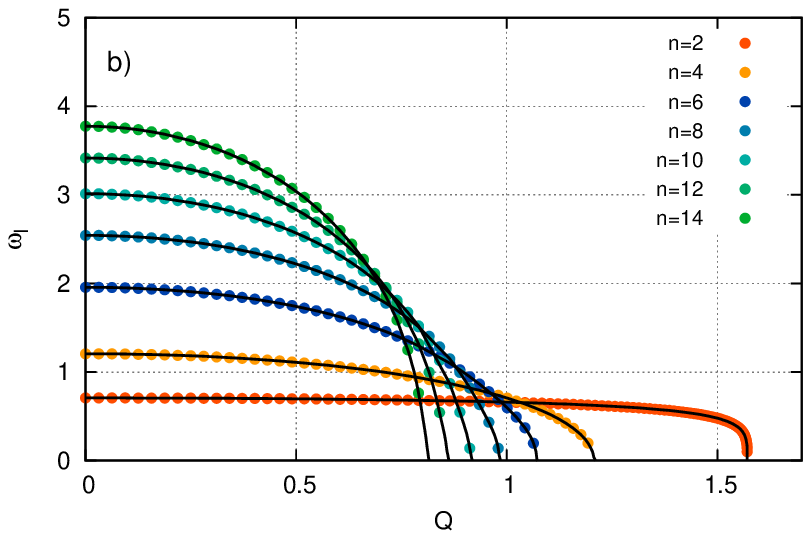}}
  \centerline{\includegraphics [width = 80mm]{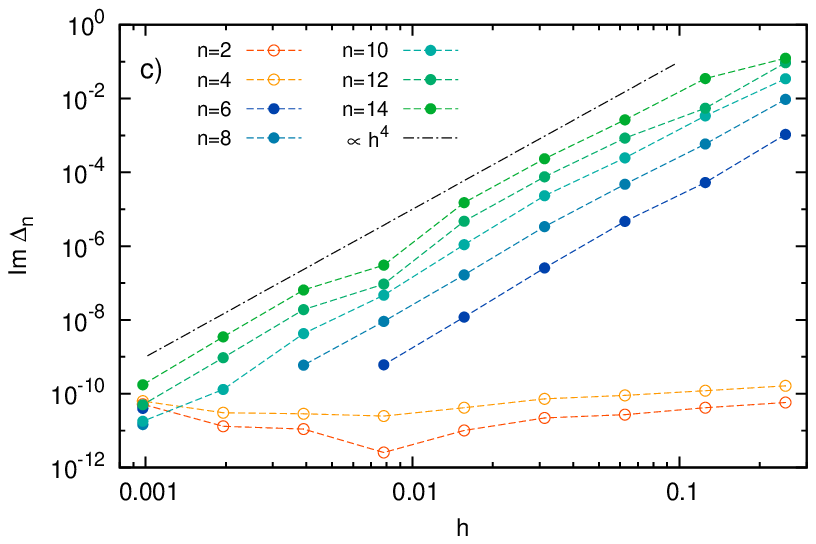}  \includegraphics [width = 80mm]{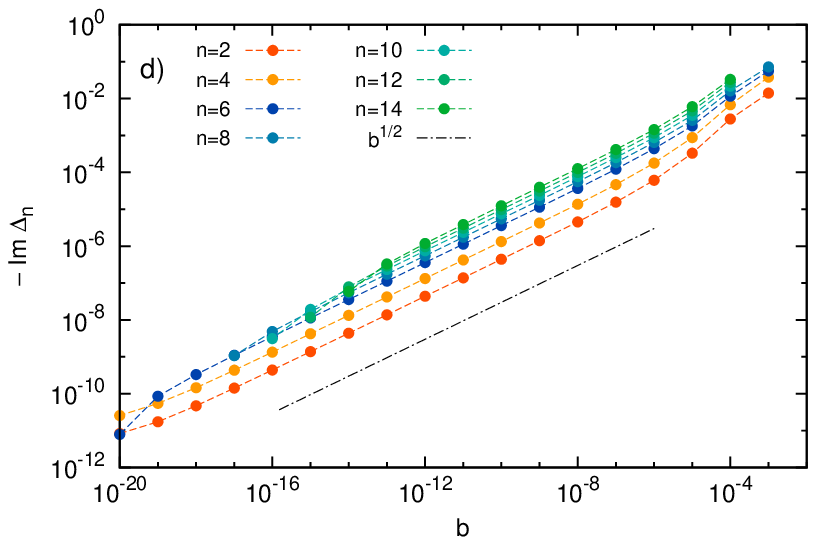}}
  \caption{Eigenvalues of 7 largest scale modes in Hunter-Shukhman model: pattern speeds (a); growth rates (b); accuracy $\Im \Delta_n$ versus element's size $h\equiv 1/N$ (c), and gravity softening $b$ (d). In the upper figures the black curves show analytic solutions of eq. (\ref{eq:dr_rigid}). Open circles in (c) for modes $n=2,4$ show $|\Delta_n|$. Dash-dotted lines in the lower panels show reference slopes.}
  \label{fig:rds}
\end{figure*}

Fig.\,\ref{fig:rds} presents eigenmodes for model parameters $\Omega=1$, $a=1$, $0\leq c_\textrm{s} \leq 1$, obtained from the matrix equation for the simplest linear radial grid with $N=256$ and $N_d=4$. Panels (a) and (b) show respectively pattern speeds and growth rates of the largest scale modes. For $Q=1$, the model has 3 unstable modes, corresponding to $n=2, 4, 6$. With pressure increasing, their growth rates decrease. The last mode saturates at $Q = \pi/2$. With pressure decreasing, new unstable modes with larger number of nodes appear. The numerical eigenvalues agree very well with analytic ones obtained from the characteristic equation (\ref{eq:dr_rigid}).

Lower panels of Fig.\,\ref{fig:rds} show accuracy of the method for $c_*=0.1$. Deviation $\Delta_n$ is measured as the difference between calculated $\omega_n$, and exact frequencies $\overline \omega_n$. Since imaginary parts of $\Delta_n$ dominate, we use it as a measure of accuracy. Panel (c) shows the decrease of Im $\Delta_n$ on parameter $h\equiv 1/N$ for $N_d=4$, and $b=10^{-20}$. It remains positive until the limiting level $\lesssim 10^{-10}$. Notice the first two modes giving $|\Delta_n|$ below this level for any $h$, naturally because the basis functions can reproduce the polynomial eigenfunctions exactly independent of the number of elements $N$. For modes $n \geq 6$, the  deviations scale roughly like $h^s$, where index $s$ is equal to 1.7, 2, 4, and 4, for $N_\textrm{d}=1,2,3,4$, respectively. 

In WKB approximation, the gravity softening parameter attenuates the force by a factor $\exp(-b k) \approx (1-b k)$, where $k$ is some radial wavenumber. The errors in frequencies thus should be expected to decrease $\propto b$. However, it is not the case in Fig.\,\ref{fig:rds}\,d, where the deviations are proportional to $b^{1/2}$. The reason of this slower attenuation is the presence of integrable singularity of the surface density perturbation (\ref{eq:efs_rigid}) in the limiting case $b=0$. This singularity requires additional regularisation at the disc edge. Namely, instead of the integrals (\ref{eq:iw}) one has to regularise 
\begin{equation}
\int\limits_{R_{n}}^{R_{n+1}} dR' \frac{w_b(R-R')}{[1-{R'}^2/a^2]^{1/2}} f(R')\ .
\label{eq:iw2}
\end{equation}
Two singularities can be treated separately unless $R$ is close to the edge. Denoting $r = a-R$, and changing variable of integration $R' = a(1-z^2)$, (\ref{eq:iw2}) can be reduced to an integral with the weight function
\begin{equation}
w'_b(z) = \ln[ (a z^2-r)^2 + b^2 ]\ . 
\label{eq:wf2}
\end{equation}
The quadrature weights for this function can be expressed through the weights $W^j_b$ of (\ref{eq:iw}) as follows:
\begin{equation}
{W'}^j_b \equiv W^j_{[\i b a -ra]^{1/2}} + \textrm{c.c.}\ , 
\label{eq:wg2}
\end{equation}
where c.c. stands for the complex conjugate of the preceding term. Now consider the case when both singularities coincide, i.e. $r=0$. Provided that weights $W^j_b$ attenuate the force $\propto b$, ${W'}^j_b$ will attenuate the force $\propto b^{1/2}$. A consequence of this is seen in panel (d). 

\begin{figure}
  \centerline{\includegraphics [width = 80mm]{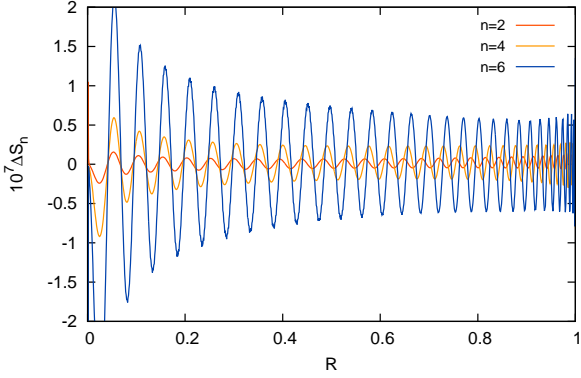}}
  \caption{Hunter-Shukhman model: deviation $\Delta S_n = \tilde \Sigma_n(R) - P_n^m(\xi)$ of 3 largest scale eigenfunctions from the Legendre polynomials.}
  \label{fig:hs_ef}
\end{figure}

The method reproduces eigenfunctions quite well, but numerical instability may occur for higher modes. Fig.\,\ref{fig:hs_ef} shows examples of the numerical instability with low amplitudes for the first 3 modes. The wave number of these oscillations is close to the short wave branch obeying WKB equation 
\begin{equation}
\kappa^2 -2 \pi G \Sigma_0 k + k^2 c^2_s = 0\ . 
\label{eq:wkb}
\end{equation}
For $Q>1$ numerical instability, as well as higher unstable modes, disappear. 

\section{The cored exponential model}

Our main objective is the calculation of unstable modes in razor-thin discs that are extensively used in the context of galactic dynamics \citep{FP84, BT08, B14}. Here we adopt the stellar dynamical cored exponential model with the soft-centred logarithmic potential
\begin{equation}
  \Phi_0 (R) = v_0 ^ 2\ln\sqrt{1 + R^ 2/R_C ^ 2}\ ,
 \label{vc_cem}
\end{equation}
where $R_C$ is the scale radius of the potential. The surface density of the disc is
\begin{equation}
 \Sigma_D (R) =\Sigma_s\exp\Big [-\lambda\sqrt{1 + R^ 2/R_C ^ 2}\Big]\ ,\quad\lambda \equiv \frac{R_C}{R_D}\ ,
 \label{exp_sd}
\end{equation}
with $ R_D $ being the disc radial scale. The distribution function of the stellar model has a free integer parameter $\overline N$ \citep[e.g.,][]{PJ15}, which controls the radial velocity dispersion (roughly $\sigma_R \approx v_\textrm{0}/(2 \overline N)^{1/2}$ in the centre). 

\begin{figure}
  \centerline{\includegraphics [width = 80mm]{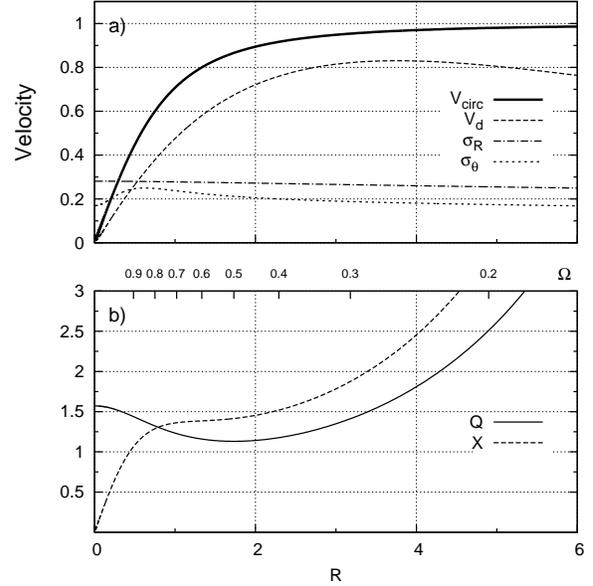}}
  \caption{The cored exponential model: (a) circular velocity profile $V_\textrm{circ}$, contribution of the disc $V_\textrm{d}$ to $V_\textrm{circ}$, radial and azimuthal velocity dispersions $\sigma_R$ and $\sigma_\theta$; (b) Toomre stability parameter $Q$ for sound speed $c_\textrm{s}(R) = \sigma_R(R)$ (solid black), and $X$-profile (see the main text). The upper scale of panel b) shows corresponding values of $\Omega(R)$; $\Omega(0) = 1$.}
  \label{fig:ced_vr}
\end{figure}

Fig.\,\ref{fig:ced_vr}\,a shows the circular velocity $V_\textrm{circ}(R)$, and the radial and azimuthal velocity dispersion profiles for $\overline N=6$, $\lambda=0.625$, $\alpha\equiv G\Sigma_s R_D/v_0^2=0.34$, in units $G = v_0 = R_C = 1$. For these parameters, the disc is submaximal as follows from  $V_\textrm{d}(R)$ profile of disc contribution to the full circular velocity.  

Our default gaseous model adopts the slowly decreasing radial velocity dispersion shown in Fig.\,\ref{fig:ced_vr}\,a for the sound speed, i.e. $c_\textrm{s}(R) = \sigma_R(R)$. Toomre stability parameter $Q = \kappa c_\textrm{s}/(\pi G\Sigma_0)$ is shown in Fig.\,\ref{fig:ced_vr}\,b. Minimum of $Q(R)$ is $Q_\textrm{min} \approx 1.13$ attained at $R\approx 1.9$. This panel also includes parameter $X$ that characterises the accuracy of the tight-winding approximation for the given disc,
\begin{equation}
  X(R) \equiv \frac{k_{\textrm{crit}}R}m = \frac{\kappa^2R}{2\pi G\Sigma_0 m}\ .
 \label{eq:X}
\end{equation}

In the following subsections, we examine this model in two cases. The first one has large softened gravity to imitate pressure while the genuine pressure term is set to zero. In the other one, we consider usual selfgravitating disc with flat pressure.

\subsection{The softened gravity disc}

\begin{figure*} 
  \centerline{\includegraphics [width = 80mm]{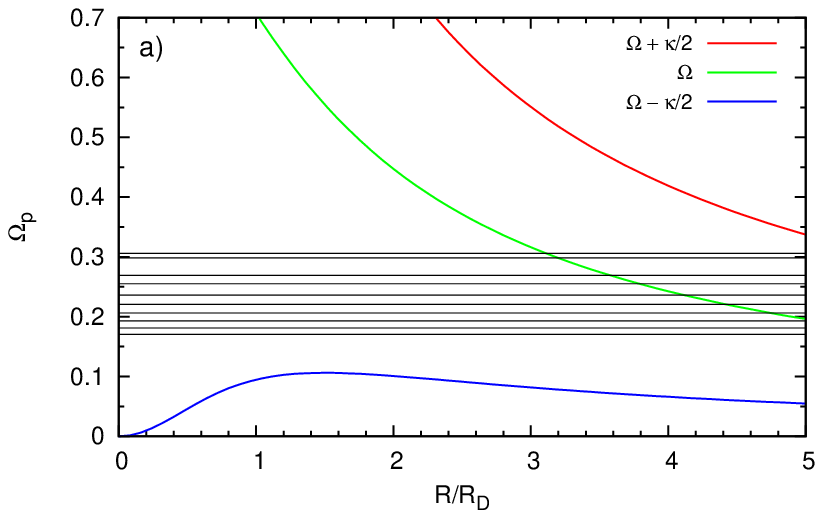} \hspace{6.5mm}\includegraphics [width = 72mm]{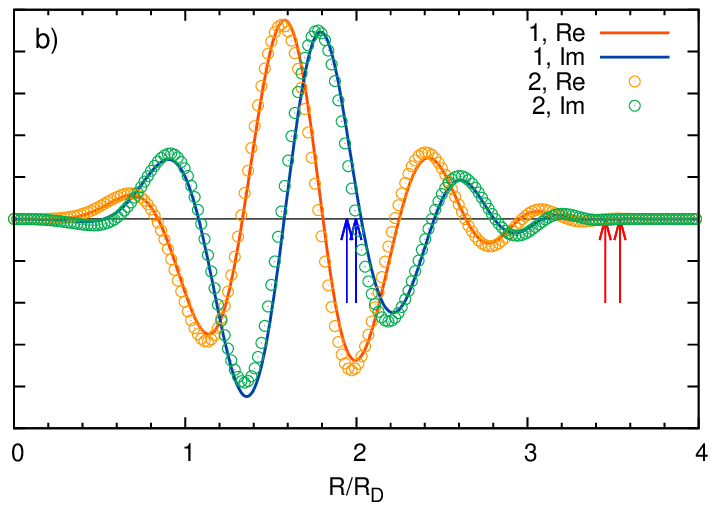}}
  \centerline{\includegraphics [width = 80mm]{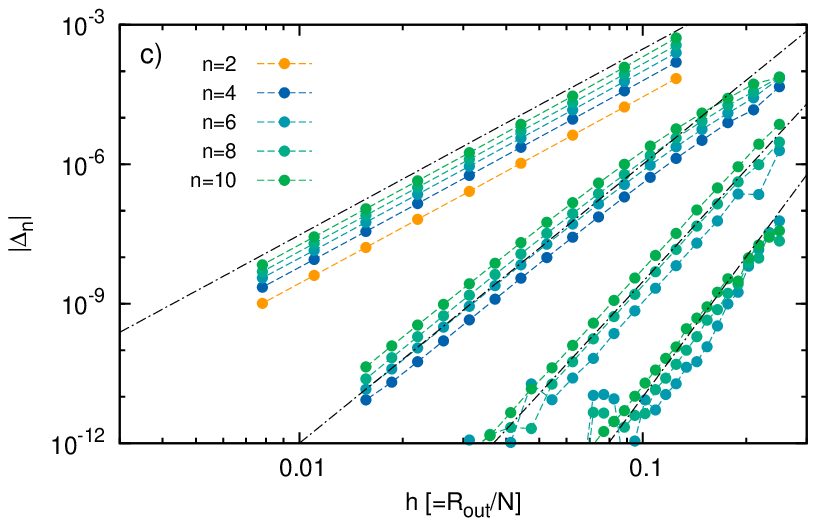} \includegraphics [width = 80mm]{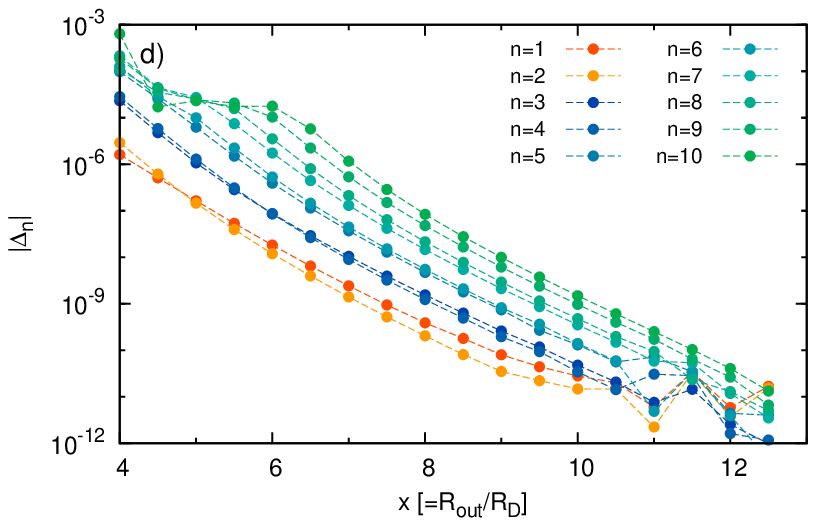}}
  \caption{The softened disc with softening $b = 0.3122$: (a) pattern speeds of the first 10 unstable modes (horizontal lines) compared to the principal frequency curves $\Omega(R)$, $\Omega \pm \kappa/2$; (b) eigenfunctions $\Sigma(R)$ (real and imaginary parts) of the first two modes; (c) accuracy $|\Delta_n|$ versus element's size $h$ for $N_\textrm{d}=1,2,3,4$; (d) accuracy $|\Delta_n|$ versus the outer boundary $R_\textrm{out}$ for $N_\textrm{d}=4$. Arrows in (b) show corotations and OLRs. Dash-dotted lines in (c) show $|\Delta_n| \propto h^s$ with $s=4,6,8,10$.}
  \label{fig:ced_soft}
\end{figure*}

It is well known that cold stellar and gaseous discs are violently unstable. The instability can be tamed by including radial velocity dispersion to stellar motion or pressure to gas. The same effect is achieved by introducing the softened gravity \citep{M71, E74, T77}. From the physical point of view, this corresponds to replacement of the point masses to little Kuzmin discs of size $b/2$. Wave propagation in the softened discs is more in line with one in the stellar discs: the short wavelength branch does not intersect the Lindblad resonances, and spiral perturbations are localised. The cored exponential disc requires $b = 0.3122$ to become stable with respect to axisymmetric perturbations $m=0$, so this value of the softened gravity roughly corresponds to $Q_\textrm{min} = 1$ for ordinary discs. Since pressure is no longer required for stabilisation, we set it to zero, and thus remove the possibility of sound waves.

Fig.\,\ref{fig:ced_soft} contains results for the softened model. Panel (a) gives pattern speeds of first 10 modes in comparison with the principle frequency curves $\Omega(R)$, $\Omega \pm \kappa/2$. None of the unstable modes have ILR. Notice that the first two pattern speeds, 0.3059 and 0.2986, nearly blend together, and the next two also appear in a pair. The effect is even more clear for lighter discs: four pairs at $\alpha = 0.25$, and five pairs at $\alpha = 0.2$ can be detected. This seems to suggest the presence of degeneracy specific to the cored exponential disc and the cored potential. No mode's pairing is observed when the disc is changed to the usual exponent, or the potential is changed to $\Phi_0(R) = v_0^2 \ln R$. 

Panel (b) shows similarity of the eigenfunctions $\Sigma(R)$ of the first paired modes. Arrows point to the positions of the corotation and the outer Lindblad resonances (OLRs). The eigenfunctions are localised within the OLRs, as expected. 

Panel (c) demonstrates convergence of FEM solutions over element's size $h\equiv R_\textrm{out}/N$. Four series of the solutions correspond to basis functions of different degree of smoothness $N_\textrm{d}$. To quantify the convergence, we derive the deviation between calculated and the reference frequencies. The latter are obtained for parameters of $N=256$ and $N_\textrm{d}=4$ and used for all series. Contrary to the model of Section 3, here real and imaginary parts of the deviation are comparable, thus we measure accuracy as the absolute value of the deviation. 

All modes in each series follows the same dependence $\Delta_n(h) \propto h^s$ given on the panel by the dash-dotted lines. The lines' slopes obey the simple dependence:
\begin{equation}
  s = 2(N_\textrm{d} +1)\ .
 \label{eq:deps}
\end{equation}
Absolute values of $\Delta_n$ increase with the mode's number $n$.

Since the eigenfunctions are localised within OLRs, the eigenmodes should not depend on $R_\textrm{out}$ unless the OLRs exceed $R_\textrm{out}$. This is confirmed by panel (d) where convergence over this parameter is shown. The deviations are obtained here for $N_\textrm{d}=4$ and fixed $h=1/32$. From the graph one can see that frequencies of the first 4 modes obtained at our default $R_\textrm{out} = 10 R_\textrm{D}$ have accuracies $|\Delta_n| \lesssim 10^{-10}$. The same modes can be obtained with accuracy better than $10^{-5}$ already for $R_\textrm{out} = 5 R_\textrm{D}$. Irregularity seen in the upper left corner for higher modes is due to the distant OLRs that exceed $4 R_\textrm{D}$.

A. Kalnajs kindly provided first 10 most unstable modes for the model with $\alpha = 4/5\pi$, $b=0.2228$ obtained by a stellar dynamical code with accuracy better than $10^{-8}$. The softened stellar and gaseous discs are equivalent, if there are no radial velocity dispersion or pressure, so FEM should reproduce them exactly. FEM reproduced all the provided digids successfully and showed no extraneous solutions owing to improper boundary conditions.

\subsection{The gaseous disc}

\begin{figure*}
  \centerline{\includegraphics [width = 80mm]{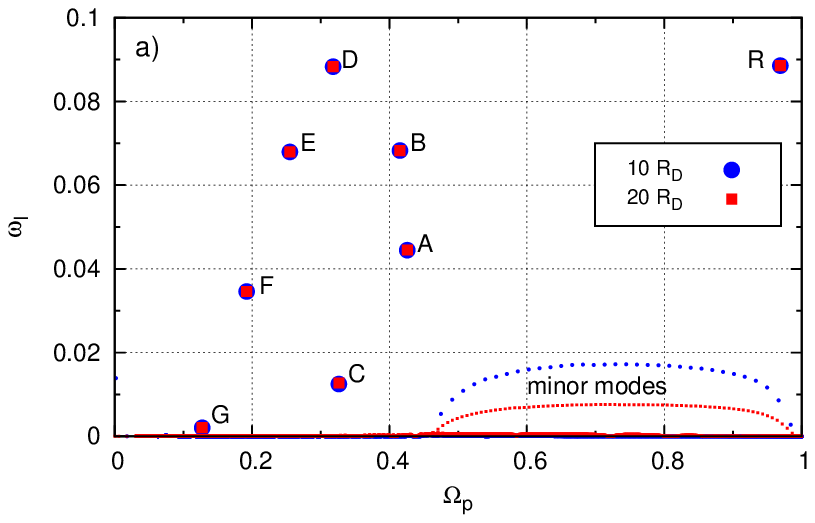} \vspace{2mm}\includegraphics [width = 80mm]{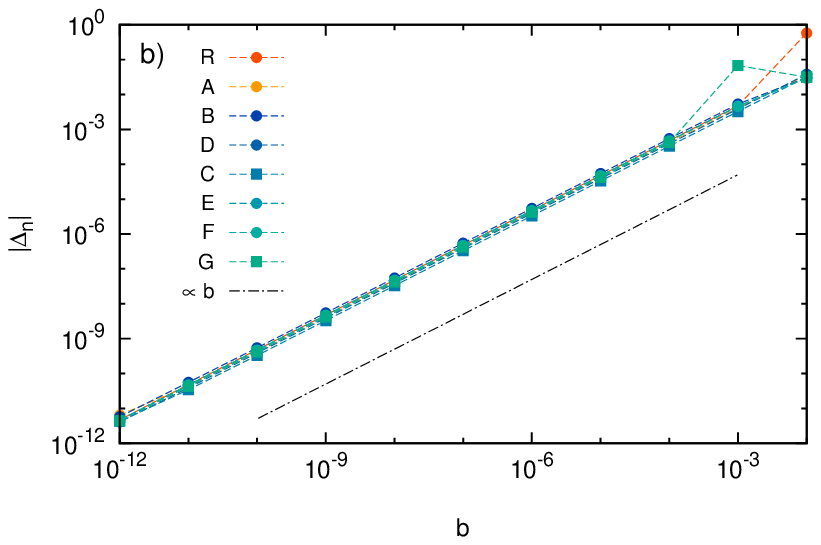} }
  \centerline{\includegraphics [width = 80mm]{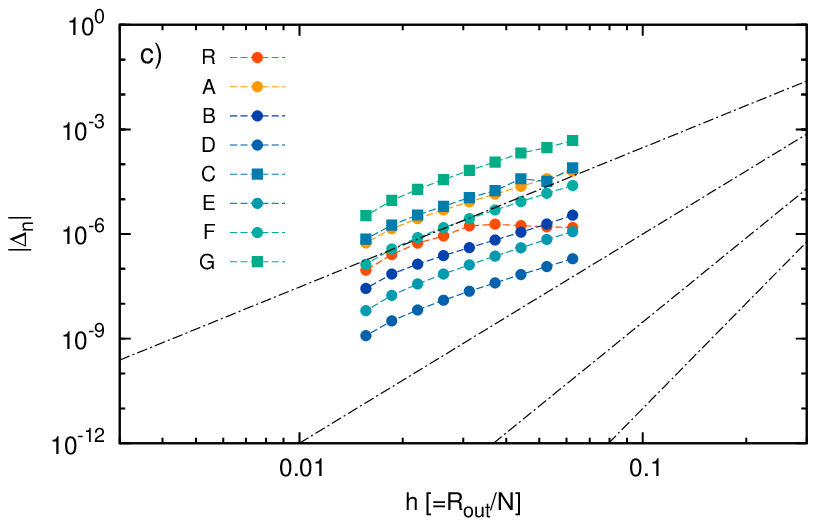}   \includegraphics [width = 80mm]{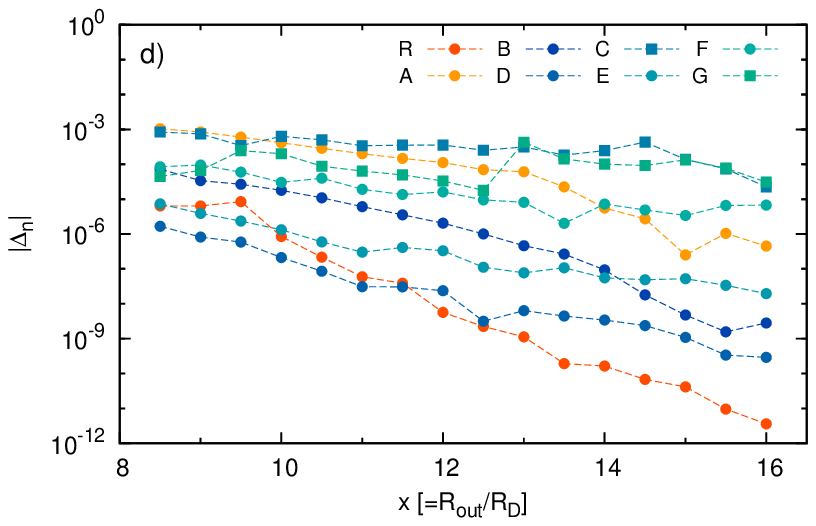}}
  \caption{The gaseous disc with $Q_\textrm{min} = 1$: (a) some `major' modes (labelled by R, A, ..., G) and a family of `minor' modes in the complex frequency plane for $R_\textrm{out} = 10 R_D$ and $20 R_D$; (b--d) accuracy $|\Delta_n|$ of the major modes versus softened gravity $b$, element's size $h$, and the outer boundary $R_\textrm{out}$ obtained for $N_\textrm{d}=4$. Dash-dotted line in (b) shows $|\Delta_n(b)| \propto b$; dash-dotted lines in (c) are the same as in Fig.\,\ref{fig:ced_soft}\,c.}
  \label{fig:ced0_Q1}
\end{figure*}

\begin{figure*} 
  \centerline{\includegraphics [width = 170mm]{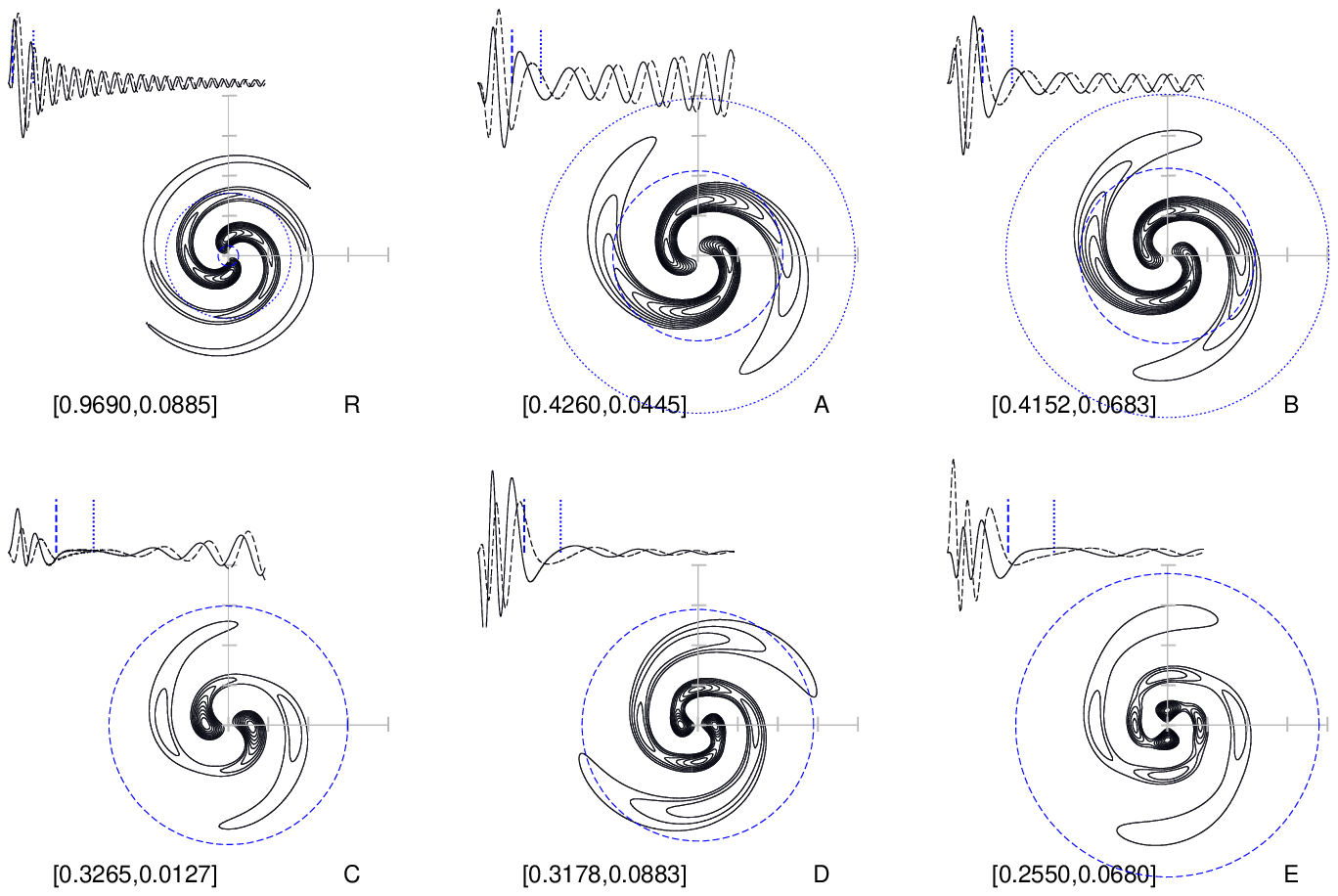}}
  \caption{Spiral patterns (isolines of $\Sigma$ corresponding to 10, 20, ... 90 per cent overdensity) for $Q_\textrm{min}=1$. Labels of the modes are given in the lower right, eigenfrequencies $[\Omega_\textrm{p}, \omega_I] $ -- in the lower left. Circles in the upper row show CR and OLR, in the lower row -- only CR. Scales are given in units of $R_C$. Real and imaginary parts of the relative surface density are shown in the upper left; vertical lines there show positions of CR and OLR.}
  \label{fig:ced_gas_patt}
\end{figure*}

In this subsection we apply FEM to the razor-thin gaseous disc with pressure. 
The outer boundary condition is dictated by our wish to reproduce stellar trailing spirals. According to \citet{LBK}, they are maintained by an outward transport of angular momentum by means of the short trailing wave outside the corotation circle. In stellar dynamical models, these waves are perfectly absorbed at OLR. In the fluid model, however, it refracts and propagates further outward as an acoustic wave. Thus, contrary to the stellar disc, for which zero boundary condition beyond OLR is appropriate, gaseous discs require the radiation boundary condition. 

Calculating the eigenmodes in gaseous discs by integrating hydrodynamic equations, \citet{P83} introduced a fictitious radius $R_\textrm{o}$ between corotation resonance (CR) and OLR, at which the radiation condition is imposed. He restricted calculations to the finite region $R \leq R_\textrm{o}$, although surface density perturbations in the outer region $R>R_\textrm{o}$ should be taken into account because of the Poisson equation. This was done using an approximation based on the density wave theory. 

In present calculations, the outer boundary $R_\textrm{out}$ is suggested to be much further out. Since the disc surface density falls down rather quickly, it is reasonable to expect that the surface density perturbations of unstable modes decrease as well. Thus, their eigenfrequencies should not feel the outer boundary if it is moved away, say at radii $R \gtrsim 10 R_D$. Yet the relative surface density and the velocity perturbations can even be growing, so we do not require them to vanish on the outer boundary. 

We start with a detailed study of the model with $c_\textrm{s}(R) = \tau\sigma_R(R)$, $\tau = 0.88$ corresponding to $Q_\textrm{min} = 1$. In Fig.\,\ref{fig:ced0_Q1}, panel (a) shows spectra of modes in the frequency complex plane, obtained for two positions of the outer boundary, $10 R_D$ and $20 R_D$. There are two distinct groups of modes. The first one (called `major') consists of separate modes labelled by R, A, ... G, and marked by large filled circles and squares. These modes remain unchanged with variation of $R_\textrm{out}$. Pattern speeds of modes A and B, and C and D are paired, similar to the softened disc considered previously.

The other group is labelled as `minor modes'. Their growth rates vary roughly inversely proportional to $R_\textrm{out}$. Notice that all modes do not depend on node's number $N$, and there are no singular modes similar to van Kampen modes of the stellar discs.

In Section 3, we argued that softening should attenuate force and frequency $\propto b$. This is confirmed by direct calculations presented in panel (b). 

Panel (c) gives accuracy dependence versus the element's size $h = R_\textrm{out}/N$, evaluated for the default $R_\textrm{out} = 10 R_D$, $b=10^{-15}$. It is seen that $|\Delta_n(h)|$ do not follow any simple power law, contrary to the previous models. The best grid resolution $h=0.016$ provides accuracy better than $10^{-5}$.

Panel (d) shows convergence of the major modes when the outer boundary radius $R_\textrm{out}$ increases. For the default $R_\textrm{out} = 10 R_D$, the deviation varies from $10^{-3}$ to $10^{-6}$. In general, the higher the growth rate of the mode, the deviation is smaller and decreases faster with the outer radius.

In the preceeding sections, we used the simplest linearly spaced nodes everywhere except Fig.\,\ref{fig:ced0_Q1}\,a. The spectrum of this panel and all further results are obtained with the increasing node spacing:
\begin{equation}
	R_j = R_* \left[ \e^{q\tau_j} - 1\right]\ ,
	\label{eq:ced_grid}
\end{equation}
where $\tau_j$ are equally spaced between 0 and 1, and parameters $R_*$ and $q$ are chosen so that $R_N = R_\textrm{out}$ and $R_2-R_1 = R_\textrm{out}/6N$. This allows to increase resolution in the central part, where corotation resonances of the modes are located.

All modes (major and minor) are trailing spirals. Spiral patterns in Fig.\,\ref{fig:ced_gas_patt} show isolines of the perturbed surface density $\Sigma$, while oscillating curves to the upper left from the patterns show real and imaginary parts of the relative surface density $\eta$. Beyond OLR, the wave number $k$ of the oscillations obeys rather well the WKB square dispersion relation:
\begin{equation}
	\omega_*^2 = \kappa^2 -2\pi G \Sigma_0(R) |k| + k^2 c_\textrm{s}^2\ ,
\end{equation}
which gives acoustic waves in the region where the axisymmetric surface density $\Sigma_0(R)$ is small. The relation admits two solutions, $\pm |k(R)|$, corresponding to trailing and leading spirals. In the inner region, more complicated cubic dispersion relation is available \citep{B89}. The latter predicts a forbidden zone between corotation and OLR, which is clearly seen in the case of modes C, D, E, and can be distinguished also in A and B. A characteristic feature of R and minor modes is absence of the forbidden zone.

Amplitudes of perturbations $v_R$, $v_\theta$, $\eta$ grow towards the outer boundary in minor modes, which indicates admixture of the leading spirals. This feature can also be seen in major modes (A and C modes at $Q_\textrm{min}=1$). However, since the latter do not feel variation of $R_\textrm{out}$, the acoustic zone beyond OLR seems to be unimportant for A and C modes. 

\begin{figure*} 
  \centerline{\includegraphics [width = 70mm]{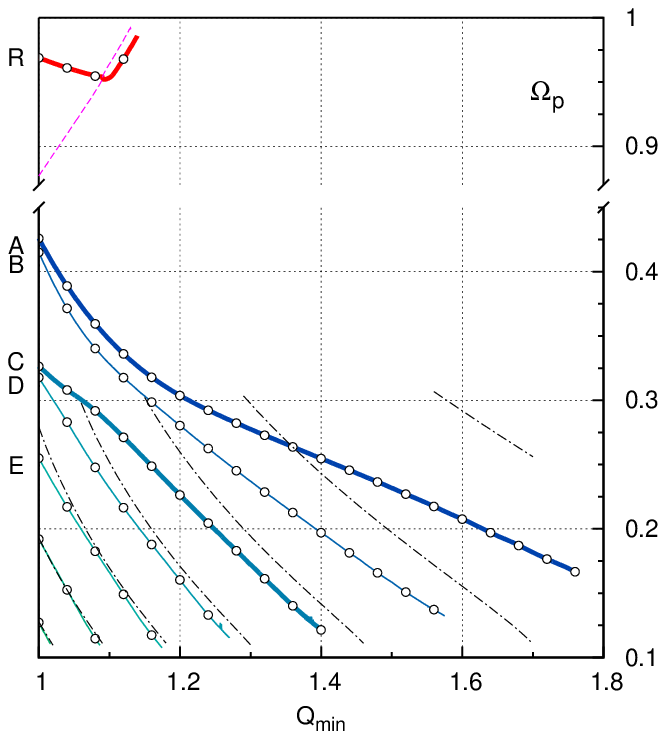} \includegraphics [width = 70mm]{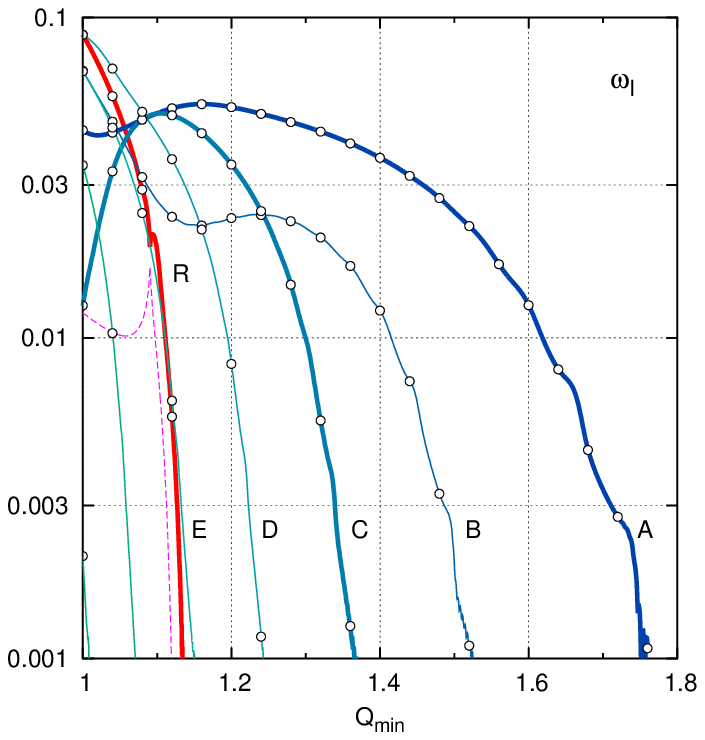}}
  \caption{Spectrum of unstable separate modes of the gaseous disc as function of $Q_\textrm{min}$. Left panel shows pattern speeds $\Omega_\textrm{p}$ inferred by matrix method and WKB solutions based on the cubic dispersion relation (dash-dotted lines). Right panel shows the major modes' growth rates. The number of elements is $N=800$.}
  \label{fig:ced_gas_spec}
\end{figure*}

While A, B, ... are usual global spiral modes, appearance of the minor modes isn't yet clear. There are two possible explanations. First, pointed out by A.\,Kalnajs, is related to ability of the cored exponential disc to emit angular momentum on corotation, which is the case when the sign of the gradient of the function
\begin{equation}
	\frac{\Omega(R)\Sigma_0(R)}{\kappa^2(R)} 
\end{equation}
is positive \citep{LBK}. The gradient is zero at $R=1.918$, and is positive within this radius. This means that modes with pattern speeds higher than $\Omega_\textrm{p}=0.462$ will emit angular momentum at corotation. This puts mode R and all minor modes in the class of angular momentum emitters on corotation. 

Another explanation -- over-reflection -- is suggested by the inverse proportionality of the growth rate $\omega_\textrm{I}$ and $R_\textrm{out}$, and the mentioned above weak sound speed dependence on radius. The propagation diagrams for minor modes include a leading branch from the outer boundary to corotation, and a trailing branch from the corotation to the outer boundary. Over-reflection takes place during transformation of the wave packet from leading to trailing \citep{GLB}, but now in the outer disc. The growth rate is inversely proportional to time needed for the wave packet to travel from the amplifier (at $R_\textrm{CR}$) to the boundary and back, 
\begin{equation}
	\omega^{-1}_\textrm{I} \propto \il_{R_\textrm{CR}}^{R_\textrm{out}} \frac{\d r}{|c_g|}\ ,
\end{equation}
where $c_g = \p \omega/\p k$ is the group velocity of the wave packet; $c_g \approx c_\textrm{s}$ beyond several radial scales of the disc. Taking into account dependence of the group velocity on $k$ and decrease of the sound speed with radius, one has 2.33 for the ratio of the travel times to the given radii, while the growth rate drops by factor 2.28. Besides, separation of the modes, $\Delta\Omega_\textrm{p}$, drops from 0.0312 to 0.0139 (ratio is 2.31), which is a consequence of a simple relation between the wavelength and $R_\textrm{out}$, $n\lambda = 2R_\textrm{out}$, where $n$ is a large integer. 
These findings suggest that minor modes obey the reflection boundary condition in the integro-differential problem.

Another support of the over-reflection conjecture can be found in Fig.\,\ref{fig:ced_vr}\,b, where the ratio of unwrapped vs. critical wavelengths 
\begin{equation}
	X \equiv \lambda_\theta/\lambda_\textrm{crit}\
\end{equation}
is given. Here $\lambda_\theta = 2\pi/k_\theta = 2\pi R/m$, $\lambda_\textrm{crit} = 4\pi^2 G \Sigma_0/\kappa^2$. The over-reflection peaks at $X \approx 1.4$ \citep[see, e.g.,][]{T81}, and is over for $X\gtrsim 3$. The pattern speeds corotating on the given radius can be found in the upper axis to the panel. The minor modes are localised in the pattern speed range corresponding to the strongest over-reflection.

A shortcoming of the method presumably connected to the insufficient resolution is a dependence of the minor modes on the smoothness of the basis functions, $N_\textrm{d}$. In particular, panel (a) calculated for $N_\textrm{d}=1$ shows 33 modes for $R_\textrm{out} =10 R_D$, and 76 mode for $R_\textrm{out}=20 R_D$. For $N_\textrm{d}=4$ only 25 modes were found for $R_\textrm{out}=10 R_D$, and 61 -- for $R_\textrm{out}=20 R_D$; the growth rates increase accordingly.
 
Further, we analyse how unstable modes evolve with pressure increasing, considering a family of models obtained from the default one by rescaling the sound speed, $c_\textrm{s}(R) = \tau\sigma_R(R)$. The scaling factor $\tau$ varies in the range $[0.88, 1.59]$, resulting in variation of $Q_\textrm{min}$ from 1.0 to 1.8. Fig.\,\ref{fig:ced_gas_spec} presents patterns speeds and growth rates of the major eigenmodes. For this disc, the highest $\Omega_\textrm{p}$ that would admin an ILR is 0.106. The pattern speeds always lie above this value, and decrease monotonically with increasing pressure. They can be compared with WKB theory based on cubic dispersion relation $D(\omega, k, R)$, which can grasp transformation of short leading to short trailing spirals \citep{B89}. The WKB solutions satisfying Bohr-Sommerfeld quantum condition
\begin{equation}
	\oint k(\omega_\textrm{R}, R) \d R = (2n+1)\pi\ ,\quad n=0,\,1\, ...
\end{equation}
are shown in the left panel by black dot-dashed curves (the upper right curve corresponds to $n=0$). Good agreement is seen only for $n\geq 5$ (modes E, F, G).

Spiral pattern of mode A has two maxima at the left end of the track, $Q_\textrm{min} = 1$, but then transforms into 1-maximun pattern at $Q_\textrm{min} \approx 1.22$. At slightly larger $Q_\textrm{min}$, mode B undergo transformation from 1 to 2-maxima mode, while mode C -- from 2 to 3-maxima mode at $Q_\textrm{min} \approx 1.16$. Modes D, E, F, G preserve number of maxima (3, 4, 5, 6, respectively) in the whole range of $Q_\textrm{min}$.

The right panel of Fig.\,\ref{fig:ced_gas_spec} shows evolution of the growth rates $\omega_\textrm{I}$ of the major modes. Modes A, B, C exhibit interesting nonmonotonic behaviour at the left end, but decrease monotonously after $Q_\textrm{min} \approx 1.2$. Mode A determines the instability properties of the disc for nearly the whole range of $Q_\textrm{min}$, and saturates the latest at $Q_\textrm{min} \approx 1.76$. This is in good agreement with the stability criterium given by \citet{PPS97a}, who predict that gaseous discs with flat rotation curves are completely stable to all nonaxisymmetric perturbations if $Q > \sqrt{3}$.

\section{Conclusions}

In the paper we offer a simple method for calculating eigenmodes (frequencies and spiral patterns) of razor-thin gaseous selfgravitating discs, in which the traditional integration of hydrodynamic equations is replaced by a matrix linear eigenvalue problem. The matrix equation is a so-called weak form of the well-known equations describing the gaseous discs, so it in no way relies to the asymptotic WKB theory. An obvious advantage of the method is that a whole spectrum of unstable modes can be obtained at once. Besides, the matrix elements can be computed in parallel using GPU technologies that potentially makes study of gaseous discs extremely fast and accurate.

While different boundary conditions can be applied in the traditional scheme, it appears that this matrix method is not so flexible. In fact, one can demand only that each of three unknown functions (surface density and velocity components) takes zero or non-zero value. In principle, it can lead to spurious solutions, which need to be eliminated by examining the eigenfunctions. But in the models considered here we obtained only physically reasonable unstable modes. 

The method is tested on three models. The first one is an exactly solvable model with the uniform rotation and polytropic index $\gamma=3$. This is a relatively simple model with polynomial eigenfunctions, except that it needs a special care when regularizing the Green's function singularity on the disc edge.
Use of smooth basis functions, $N_\textrm{d} \geq 2$ is an advantage, as it allows to use smaller number of elements $N$ to achieve needed accuracy. At small gravity softening, the dominating error term lead to force correction law $\propto b^{1/2}$. This behaviour is unusual, and results from the presence of integrable singularity on the edge of the disc. For $Q<1$ the model is vigorously unstable, reflecting in numerous modes with high growth rates. Although errors for the higher modes obey the same dependences on $h=1/N$ and $b$ as the largest scale modes, their accuracy is poorer presumably due to the numerical instability taking place at $Q<1$.

The second and third models are based on the stellar disc with cored exponential profile in the cored gravitational potential. First of them has a finite gravity softening to eliminate axisymmetric instability and no pressure, and thus the possibility of sound waves is removed. In this limit, the gaseous disc is identical to the softened stellar disc with no radial velocity dispersion, so direct comparison of the eigenmodes is possible. A.\,Kalnajs kindly provided the stellar eigenmodes for comparison, which appeared satisfactory. A characteristic feature found in the cored exponential disc and the cored potential is a degeneracy leading to paring of unstable modes.

The last model -- gaseous disc with pressure -- turned out to be the most difficult one, because it allows sound waves to go far away. It is usually assumed that the disc has no boundary, so the spiral waves radiate off the central part of the disc. Finite element method, however, requires the outer boundary by construction, and this feature is not physically implausible. The boundary provided a group of `minor' modes, in addition to a number of usual global spiral (`major') modes. Some of the global spiral modes can be described by WKB theory fairly well, but not the most important one. 

Near to the stability boundary of the axisymmetric modes, $Q=1$, there are several modes that demonstrate abnormal increasing of growth rates with pressure increasing. The disc becomes stable with respect to bisymmetric mode, when $Q$ exceeds 1.76 everywhere. This result agrees with the previous finding of the author with collaborators that $Q=\sqrt{3}$ is a stability boundary for all $m$, providing the flat rotation curve profile \citep{PPS97a}. In this respect, gaseous discs are significantly less unstable than stellar discs that require $Q\gtrsim 3$ for stability \citep{PPS97b}.

As a next step, we plan to perform a detailed comparison of stability properties of gaseous and stellar discs for soft-centred and cuspy galactic models.

\section*{Acknowledgments}

The author thanks A.\,Kalnajs for discussions and suggestions on improving the original version of the paper, and I.\,G.\,Shukhman for pointing to the generalisation of the cold model suggested by \citet{H63}. This work was supported by the Sonderforschungsbereich SFB 881 ``The Milky Way System'' (subproject A6) of the German Research Foundation (DFG), and by the Volkswagen Foundation under the Trilateral Partnerships grant No. 90411. The author also acknowledges financial support by the Russian Basic Research Foundation, grants 15-52-12387, 16-02-00649, and by Basic Research Program OFN-15  `Interstellar and intergalactic media: active and elongated objects' of Department of Physical Sciences of RAS.

\end{document}